\documentstyle[11pt,aas2pp4]{article}
\lefthead{Walsh et al}
\righthead{Abundances of PN in Sagittarius dwarf galaxy}
\begin{document}
 
\title{Chemical Abundances of Planetary Nebulae in the 
Sagittarius Dwarf Elliptical Galaxy \footnote{Based on observations at the
European Southern Observatory, La Silla}}
\author{J. R. Walsh}
\affil{Space Telescope European Co-ordinating Facility,
European Southern Observatory,
Karl-Schwarzschild Strasse 2, 
D-85748 Garching bei M\"{u}nchen, Germany}
\authoremail{jwalsh@eso.org}
\and
\author{G. Dudziak}
\affil{European Southern Observatory,
Karl-Schwarzschild Strasse 2, 
D-85748 Garching bei M\"{u}nchen, Germany}
\authoremail{gdudziak@eso.org}
\and
\author{D. Minniti}
\affil{Lawrence Livermore National Laboratory,
MS L-413, P.O Box 808, Livermore, CA 94550}
\authoremail{dminniti@llml.gov}
\and
\author{A. A. Zijlstra}
\affil{European Southern Observatory,
Karl-Schwarzschild Strasse 2, 
D-85748 Garching bei M\"{u}nchen, Germany}
\authoremail{azijlstr@eso.org}

\begin{abstract}
Spectrophotometry and imaging of the two planetary nebulae
He~2-436 and Wray~16-423, recently discovered to be in the 
Sagittarius dwarf elliptical galaxy, are presented. Wray~16-423 
is a high excitation planetary nebula (PN) with a hot central star.
In contrast He~2-436 is a high density nebula with a cooler central
star and evidence of local
dust, the extinction exceeding that for Wray~16-423 by E$_{B-V}$ =
0.28. The extinction to Wray~16-423, (E$_{B-V}$=0.14) is consistent
with the line of sight extinction to the Sagittarius Dwarf. 
Both PN show Wolf-Rayet features in their spectra, although the lines are
weak in Wray~16-423. Images in [O~III] and H$\alpha$+[N~II],
although affected by poor seeing, yield a diameter of
1.2$''$ for Wray~16-423 after deconvolution, whilst He~2-436 was unresolved.
He~2-436 has a luminosity about twice that of Wray~16-423 and its 
size and  high density suggest a younger nebula. In
order to reconcile the differing luminosity and nebular properties
of the two nebulae with similar age progenitor stars, it is suggested 
that they are on He burning tracks

  An abundance analysis is presented for both PN 
using empirical abundance determinations. The abundance pattern 
is very similar in both nebulae and both show an oxygen
depletion of $-$0.4 dex with respect to the mean oxygen abundance of
Galactic planetary nebulae and [O/H] = $-$0.6 . The Sagittarius PN 
progenitor stars are representative of the higher metallicity tail 
of the Sagittarius population. The pattern of abundance depletion is 
similar to that in the only other planetary
nebula in a dwarf galaxy companion of the Milky Way, that in Fornax, 
for which new spectra are presented. However the abundances
are larger than for Galactic halo PN suggesting a later formation age.
The oxygen abundance of the Sagittarius galaxy
deduced from its PN, shows similarities with that of dwarf ellipticals 
around M~31, advancing the notion that this galaxy was a dwarf 
elliptical before its interaction with the Milky Way.
\end{abstract}


\section{Introduction}
The Sagittarius dwarf elliptical galaxy (synonymous with the Sagittarius dwarf
spheroidal) was discovered only in 1994 by 
Ibata et al. (1994, 1995) and lies in a region 
some ten degrees from the Galactic Centre, at a distance of 25 $\pm$ 3 kpc. 
Extensive photometry and star counts have shown
that the full extent of this galaxy is larger than first revealed.
Alard (1996) and Alcock et al. (1997) found RR Lyrae stars belonging to this
galaxy in fields out to Galactic latitude $b = -2^{\circ}$ and
Mateo et al. (1996) and Ibata et al. (1997) found members extending 
below $b=-25^{\circ}$, implying an extension of at
least $25^{\circ}$, or more than 10 kpc for a distance of 25 kpc.
The galaxy appears to be prolate with an axis ratio of 3:1 (\cite{IBA97}). 
Based on determination of its orbit, the galaxy must have undergone 
many orbital crossing of the Galaxy, but although distorted has not been
disrupted (\cite{IBA97}; \cite{MAT96}). There are four 
globular clusters associated
with the Sagitarius dwarf, of which M~54 may constitute the nucleus
(\cite{IG94}; \cite{DAC95}; \cite{MIN96}). The galaxy appears to be more 
massive than the majority of Galactic
dwarf ellipticals and its luminosity is $\geq$2$\times$10$^{7}$ L$_\odot$ 
(\cite{MAT96}). Zijlstra \& Walsh (1996) during the course of a survey of the
dynamics of planetary nebulae (PN) toward the Galactic Bulge discovered that two 
previously catalogued (Galactic) PN in the direction towards the
Sagittarius dwarf had radial velocities in agreement with that of the
tidal tail of the Sagittarius galaxy. 
Figure 1 in Zijlstra \& Walsh (1996) shows the position of these
two PN (He~2-436, (\cite{HE67}) and Wray~16-423 (\cite{WR66})) in 
relation to the galaxy and its globular clusters. A third 
PN (PRMG-1) also lies in the direction towards this galaxy ($l=6.0, b=-41.9$), 
and has been considered as a halo planetary nebula on the basis of its low 
abundances (\cite{PRMG89}). PRMG-1 clearly does not belong to
the Galactic bulge population so could be a possible member of the Sagittarius dwarf;
lacking a radial velocity measurement however, its membership is unproven. 

  In view of the importance of these PN for studying the stellar population 
and star formation history of the closest dwarf galaxy to our 
Galaxy, narrow and broad slit spectrophotometry and emission line imagery 
were obtained of both He~2-436 and Wray~16-423 in order to 
study their chemical abundances and their size and structure. In
addition, spectroscopy of the only other PN known in a dwarf elliptical
satellite galaxy of the Milky Way, that in Fornax (\cite{DAN78}), was also obtained.
Section 2 presents the observations and reductions and Section 3
the analysis. Section 4 discusses the abundance pattern revealed by these 
three nebulae in the context of the chemical evolution of the 
local dwarf galaxies.

\section{Observations and Reductions} 

\subsection{Spectrometry}
Long slit observations of both Sagittarius dwarf galaxy PN were obtained
with the ESO NTT and Nasmyth imager/spectrograph EMMI using the Red Imaging
and Low Dispersion Mode (RILD) mode, with grism \#3 and a slit width of 
1.5$''$. The RILD mode allows imaging, long-slit grism spectroscopy,
slitless and multi-object spectroscopy over the wavelength range 4000 -
10000\AA. The detector was a Tek 2048$\times$2048 thinned chip (ESO\#36, 
TK512CB) with 24$\mu$m pixels (0.27$''$) and the spectral resolution 
(2 pixels) was 4.5\AA. Details of the observations are listed in Table 1.
He-Ar comparison lamp spectra and broad slit (5.0$''$) exposures 
of the spectrophotometric standard EG274 (\cite{HA94}) were obtained
for wavelength and flux calibration respectively. 

  Both Sagittarius PN were also observed with the ESO
1.52m telescope and Boller and Chivens spectrograph with a Loral/Lesser 
2048$^{2}$ CCD (ESO \#39) as detector (15$\mu$m pixels and read-out noise 
5e$^-$). Low dispersion spectra (131\AA/mm) covering the optical range
at a resolution of 5\AA\ (up to 7\AA\ in the blue) were obtained with both 
short and long exposures to record bright emission lines unsaturated as well as
weak lines. Broad slit spectra were also obtained of Wray~16-423 for
determination of the absolute line flux.
Higher dispersion spectra with a holographic grating (\#32) 
giving a resolution of 0.99\AA\ were used to resolve
the density sensitive [O~II] 3627,3729\AA\ doublet, and optimally record the weak
C~II 4267\AA\ line. All details are given in Table 1. Broad
slit observations of LTT~3864 (\cite{HA94}) accompanied all
the PN observations for flux calibration. For the low dispersion spectra, the 
very high quantum efficiency in the blue for this coated CCD meant that
there was significant contamination of the spectra from second order light
at wavelengths above 6300\AA, despite using a GG345 blocking filter. As a result
the sensitivity curves derived from the spectrophotometric standard star spectra
are not reliable in the red ($>$6500\AA) despite using a standard star of
late spectral type. Exposures of a He-Ar lamp were used 
for the purposes of wavelength calibration.
The seeing was around 1-1.5$''$ for the 1.5m observations and the observing nights 
photometric except the last when the higher dispersion spectra were obtained.

  All the 2-D spectra were reduced in the usual way, using the
spectroscopic packages in IRAF\footnote{IRAF is distributed by the
National Optical Astronomy Observatories, which is operated by the
Associated Universities for Research in Astronomy, Inc., under
contract to the National Science Foundation.}. The bias was
subtracted, the pixel-to-pixel variations were rectified by
division by a mean sky flat field exposure and the data were rebinned into
channels of constant wavelength width by fitting third
order polynomials to the comparison lamp spectra. The atmospheric
extinction was corrected and an absolute flux calibration was applied
from the observation of the spectrophotometric standard.
Spectra were extracted over the total extent of the nebulae along
the slit (large for the NTT observations on account of the poor
focus) and the mean sky from along the slit was utilized for sky 
subtraction.

  The Fornax PN was observed with the SAAO 1.9m Cassegrain unit spectrograph
and a Reticon photon counting detector (RPCS). 1-d spectra (1024 pixels)
of object and offset sky are obtained through two apertures
separated by 30$''$ on the sky of length 6$''$. Low dispersion spectra with a 
210\AA/mm grating giving a resolution of 8\AA\ were obtained.  A 
series of six 1200s exposures was obtained alternating the object in 
each aperture. Exposures of the spectrophotometric standard star LTT~1020 
(\cite{HA94}) were obtained in a similar manner.
All the data were wavelength calibrated with Cu-Ar lamp exposures. Full
details are listed in Table 1. The spectra of the standard and the 
target were reduced separately for
each aperture. After wavelength calibration and extinction correction,
sky in the paired aperture was subtracted, the separate spectra averaged 
and the spectrum flux calibrated by the standard star spectrum in the 
same aperture. The final spectrum was formed from the averaged spectra 
from both sets of calibrated aperture spectra.

\begin{table*}
\caption{Log of spectrometric observations \label{tbl1}}
\begin{center}
\begin{tabular}{llcccrc}
Telescope/ & Object & Lines & $\lambda$ range & Slit width & Exp. & Date \\
Instrument &         & mm$^{-1}$ & (\AA)      & ($''$)     &  (s)~ & \\
\tableline

NTT/EMMI & He~2-436 & 360 & 3800-8400 & 1.5 & 1200 & 1996 May 28 \\
RILD     & Wray~16-423 & 360 & 3800-8400 & 1.5 & 1000 & 1996 May 28 \\
         & Wray~16-423 & 360 & 3800-8400 & 5.0 & 200 & 1996 May 28 \\
         &             &     &           &     &     &              \\
ESO~1.5m/ & He~2-436 & 600 & 3600-7300 & 2.0 &  600 & 1996 June 03 \\
B \& C   & He~2-436 & 600 & 3600-7300 & 2.0 & 1800 & 1996 June 03 \\
         & He~2-423 & 2400 & 3490-4490 & 2.0 & 2400 & 1996 June 06 \\
         & Wray~16-423 & 600 & 3600-7300 & 2.0 & 180 & 1996 June 04 \\
         & Wray~16-423 & 600 & 3600-7300 & 2.0 & 1800 & 1996 June 04 \\
         & Wray~16-423 & 600 & 3600-7300 & 6.4 & 900 & 1996 June 04 \\
         & Wray~16-423 & 2400 & 3490-4490 & 2.0 & 3600 & 1996 June 06 \\

         &             &     &           &     &     &            \\
SAAO 1.9m/ & Fornax PN & 300 & 3300-7500 & 1.8 & 7200 & 1990 Sept 22 \\
RPCS      &           &     &           &     &     &  \\
\end{tabular}
\end{center}
\end{table*}

\subsection{Imaging}
Accurate J2000 coordinates were determined for both confirmed Sagittarius galaxy PN,
and for the possible candidate PRMG-1, and are listed in Table 2.

\begin{table*}
\caption{Coordinates of confirmed and possible planetary nebulae in the 
Sagittarius Dwarf \label{tbl2}}
\begin{center}
\begin{tabular}{lrrrrrrc}
\hline
 ~~PN & \multicolumn{6}{c}{RA ~~~(J2000)~~~ Dec} & Heliocentric  \\
   & $h$ & $m$ & $s$ & $^\circ$ & $'$ & $''$ & velocity \\
   &     &     &     &          &     &      & (kms$^{-1}$) \\
\tableline
He~2-436 & 19 & 32 & 06.69 & -34 & 12 & 57.8 & +133 \\
Wray~16-423 & 19 & 22 & 10.52 & -31 & 30 & 38.8 & +133 \\
PRMG-1 & 21 & 05 & 53.57 & -37 & 08 & 40.7 & \\ 
\end{tabular}
\end{center}
\end{table*}

On the same night as the NTT spectroscopic observations,
[O~III]5007\AA\ and H$\alpha$+[N~II] emission line images were obtained
of both He~2-436 and Wray~16-423 with EMMI.
Table 3 presents the details of the filter observations. 
The detector was the same as that used for spectroscopy.
The NTT observations were obtained
in Director's Discretionary Time following telescope and instrument
testing. An image analysis had not been performed to set the 
active optics. The images therefore suffer from poor focus and uncorrected 
coma. The seeing was in addition not good with values in excess of 1$''$,
and the measured seeing (FWHM) of star images was 1.5$''$.
In the limited time available, no flat field frames could be taken nor
spectrophotometric calibration stars observed. The images were therefore
treated in essentially their raw form. 

\begin{table*}
\caption{Log of NTT/EMMI imaging observations \label{tbl3}}
\begin{center}
\begin{tabular}{llccrc}
\hline\noalign{\smallskip}
~Target & Filter & Central $\lambda$ & Width & Exp. & Date \\
~Name   & Name   & ~~(\AA)  & ~(\AA) & (s)~ & \\
\tableline

He~2-436 & {[O III]} & 5013 & 59 & 300 & 1996 May 27 \\
He~2-436 & H$\alpha$+{[N~II]} & 6566 & 61 & 300 & 1996 May 27 \\
Wray~16-423 & {[O III]} & 5013 & 59 & 200 & 1996 May 27 \\
Wray~16-423 & H$\alpha$+{[N~II]} & 6566 & 61 & 300 & 1996 May 27 \\
\end{tabular}
\end{center}
\end{table*}

\section{Results}

\subsection{Spectrometry}
All the spectra were analysed by interactively fitting Gaussians to the
emission lines. Errors on the line fits were computed taking into account the
photon noise on the data, the read-out noise and the underlying continuum
and sky, and errors on the lines fluxes were propagated. Within the errors
the ESO 1.5m low and intermediate dispersion data were consistent for each PN.
However the reddening for He~2-436, determined from the ratios of the observed 
Balmer lines to the Case B values, for the NTT grism data was larger than 
the value determined from the ESO 1.5m data. Also no consistent fit to a single 
value of c was determined from the H$\alpha$/H$\beta$, H$\beta$/H$\gamma$ and 
H$\beta$/H$\delta$ ratios for the NTT observations of Wray~16-423. 

The NTT data suffered from poor focus which may have affected the 
spectral transmission,
even though the observations were taken at modest zenith distances (29$^\circ$
for the standard, 26$^\circ$ for He~2-436 and 16$^\circ$ for Wray~16-423). 
The He~2-436 NTT spectrum indicated a consistent, but high,
value of the reddening from ratios of Balmer lines and was taken at similar
airmass to the spectrophotometric standard, whilst the airmass of the Wray~16-423
observation differed. The different values of reddening remain puzzling but the 
dereddened spectra from the ESO 1.5m and NTT data for He~2-423 were very 
similar tending to indicate that there was a systematic (colour) effect for the NTT 
spectra. \cite{JK93} have discussed in detail the problems encountered
in spectrophtometry of faint emission line objects such as PN.
In our case the lack of image analysis most probably resulted in the 
image being centred in the slit only for a narrow range of wavelengths
and thus the effects of differential 
atmospheric refraction were exaggerated. In particular uncorrected coma 
(which is wavelength dependent) could
account for the effects. It was decided to adopt the ESO 1.5m reddening values 
and use the relative line fluxes in the blue. Where fainter lines 
were detected in the NTT spectra, or better detections of fainter lines,
these values were used, taking the ratio to the nearest strong line; errors 
were propagated accordingly. In the
red ($\lambda$$^>_\sim$6300\AA) the second order contamination affected the 
ESO 1.5m fluxes, so the NTT fluxes were used in this region scaled to
the dereddened H$\alpha$ flux. Thus the composite spectrum was constructed 
in dereddened space and then reddened. The Galactic reddening law (\cite{SE79}) 
was used throughout. 

\begin{figure}
\epsscale{1.00}
\plotone{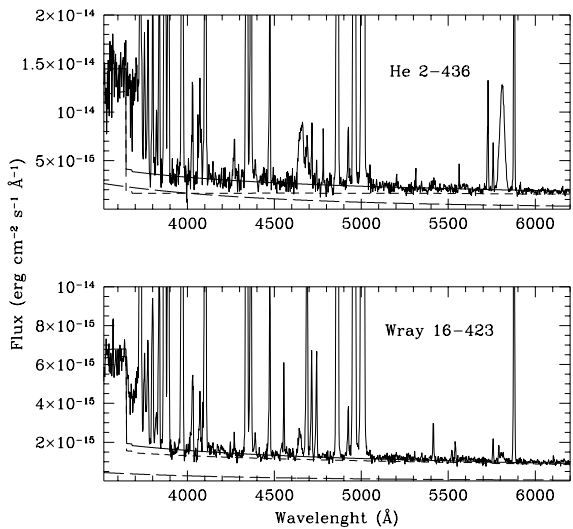}
\caption[]{(a) The dereddened spectrum (thick bold line), the nebular continuum 
(short-dashed line), the 
adopted black body flux (long-dashed line) and the sum of black body and 
nebular continuum (continuous line) are shown for the fit to the spectrum of He~2-436.
(b) The dereddened spectrum, the nebular continuum and the 
adopted black body flux are similarly shown for Wray~16-423.\label{fig1}}
\end{figure}

  Tables 4 and 5 list the resulting observed and dereddened nebular spectra 
of He~2-436 and Wray~16-423 respectively. The errors are the 
propagated 1$\sigma$ errors 
from the line fits; the sometimes differing error values for lines 
of similar strength result from the source of the
line strengths (ESO 1.5m grating low dispersion (3600-6400\AA), ESO 1.5m 
intermediate dispersion for the 3500-4500\AA\ region and NTT spectrum 
(4100-8000\AA)). Given that the final spectra in Tables 4 and 5 are composites 
of several individual spectra, there may be small systematic errors which increase
slightly the errors above those tabulated.  The errors on the reddening have not
been propagated to the dereddened relative line intensities in Tables 4 and
5. Table 6 lists the derived parameters 
from the spectra. Noteworthy is the differing value of c for the two nebulae.
He~2-436 has a much higher extinction (by 0.28 in E$_{B-V}$) than Wray~16-423.
The extinction to Wray~16-423 is the same as that for the stellar component of
the Sagittarius dwarf (\cite{IG95}; \cite{MAT95}), further corroborating 
evidence that it is indeed
a bona fide member of this galaxy. The higher extinction to He~2-423 is either 
evidence for patchy extinction along the two line of sight through
the Galaxy to this PN or suggests that there is local 
extinction in He~2-423. Given that He~2-423 is a compact, dense nebula with 
strong IR emission (it corresponds to the IRAS point source 19288-3419 with 
well detected 12 and 25$\mu$m fluxes), then there is probably a large 
component of local extinction present. 

The absolute observed H$\beta$ flux for Wray~16-423 was determined from the wide slit
observation. For He~2-436 no reliable wide slit observation was obtained
and the result from \cite{WE83} was used. The H~I (and for Wray~16-423 
the He~II) black body Zanstra temperatures were
determined from the long exposure (narrow slit) spectra after scaling the 
total spectrum (line + continuum) by the ratio of the absolute 
H$\beta$ flux to the observed H$\beta$ flux, reddening correcting the spectra
and subtracting the H and He nebula continuum (bound-free, free-free and
2-photon), using the DIPSO analysis package (\cite{HO95}). 
The dereddened stellar B magnitudes were determined from the continuum
flux and are listed in Table 6. Figure 1 shows a composite of the dereddened
spectra and the nebular and stellar continua; the stellar continuum is shown 
as a black body of temperature 60000K for He~2-436 and 90000K for Wray~16-423.

\begin{deluxetable}{llrcrc}
\tablecolumns{6}
\tablewidth{0pc}
\tablecaption{Observed and reddening corrected spectrum of He~2-436 \label{tbl4}}
\tablehead{
\colhead{$\lambda$ (\AA)} & \colhead{IDENTIFICATION}  & \multicolumn{2}{c}{OBSERVED} &
\multicolumn{2}{c}{DEREDDENED} \\
\colhead{}    & \colhead{} & \colhead{FLUX}  & \colhead{ERROR} & \colhead{FLUX} & 
\colhead{ERROR}
} 
\startdata     
 3704 & H I (B16)        &   2.70 & 0.90 &   3.90 & 1.29 \nl 
 3712 & H I (B15)        &   2.97 & 0.66 &   4.28 & 0.95 \nl
 3722 & H I (B14)        &   3.08 & 0.48 &   4.42 & 0.69 \nl
 3726 & [O II]           &   5.36 & 0.44 &   7.68 & 0.63 \nl
 3728 & [O II]           &   2.18 & 0.61 &   3.13 & 0.87 \nl
 3734 & H I (B13)        &   2.06 & 0.46 &   2.95 & 0.66 \nl 
 3749 & H I (B12)        &   4.54 & 0.70 &   6.47 & 0.99 \nl 
 3770 & H I (B11)        &   3.72 & 0.75 &   5.28 & 1.07 \nl
 3798 & H I (B10)        &   3.99 & 0.57 &   5.62 & 0.81 \nl
 3820 & He I             &   0.86 & 0.37 &   1.21 & 0.51 \nl
 3836 & H I (B9)         &   5.57 & 0.34 &   7.76 & 0.47 \nl
 3869 & [Ne III]         &  41.75 & 0.75 &  57.70 & 1.04 \nl 
 3889 & H I (B8) + He I  &  13.37 & 0.38 &  18.37 & 0.52 \nl
 3967 & [Ne III]         &  13.92 & 0.49 &  18.71 & 0.66 \nl
 3970 & H$\epsilon$ (B7)   &  11.97 & 0.45 &  16.08 & 0.60 \nl 
 4026 & He I             &   2.65 & 0.35 &   3.50 & 0.46 \nl 
 4068 & [S II]           &   2.13 & 0.44 &   2.78 & 0.57 \nl
 4076 & [S II]           &   0.75 & 0.34 &   0.98 & 0.44 \nl
 4101 & H$\delta$ (B6)     &  20.26 & 0.44 &  26.15 & 0.56 \nl 
 4267 & C II             &   0.79 & 0.19 &   0.97 & 0.23 \nl 
 4340 & H$\gamma$ (B5)     &  39.17 & 0.59 &  46.79 & 0.70 \nl 
 4363 & [O III]          &  12.07 & 0.24 &  14.31 & 0.29 \nl 
 4471 & He I             &   5.34 & 0.19 &   6.11 & 0.21 \nl
 4712 & [Ar IV] + He I   &   1.49 & 0.18 &   1.57 & 0.19 \nl 
 4740 & [Ar IV]          &   0.71 & 0.13 &   0.74 & 0.14 \nl 
 4861 & H$\beta$ (B4)      & 100.00 & 0.00 & 100.00 & 0.00 \nl 
 4921 & He I             &   1.67 & 0.13 &   1.63 & 0.13 \nl 
 4959 & [O III]          & 281.03 & 1.32 & 271.64 & 1.28 \nl 
 5007 & [O III]          & 831.72 & 3.45 & 790.83 & 3.28 \nl 
 5517 & [Cl III]         &   0.26 & 0.11 &   0.21 & 0.09 \nl
 5537 & [Cl III]         &   0.17 & 0.08 &   0.13 & 0.07 \nl
 5755 & [N II]           &   1.44 & 0.12 &   1.10 & 0.09 \nl
 5876 & He I             &  24.62 & 0.22 &  18.21 & 0.16 \nl 
 6301 & [O I]            &   7.28 & 0.07 &   4.90 & 0.05 \nl 
 6312 & [S III]          &   3.68 & 0.05 &   2.47 & 0.03 \nl
 6364 & [O I]            &   2.71 & 0.06 &   1.80 & 0.04 \nl
 6548 & [N II]           &   5.54 & 0.71 &   3.54 & 0.45 \nl 
 6563 & H$\alpha$ (B3)     & 442.54 & 1.49 & 282.16 & 0.95 \nl 
 6583 & [N II]           &  18.00 & 0.54 &  11.43 & 0.34 \nl 
 6678 & He I             &   7.17 & 0.08 &   4.47 & 0.05 \nl 
 6717 & [S II]           &   0.79 & 0.05 &   0.49 & 0.03 \nl
 6731 & [S II]           &   1.79 & 0.04 &   1.11 & 0.03 \nl
 7065 & He I             &  22.13 & 0.34 &  12.84 & 0.20 \nl
 7136 & [Ar III]         &  11.31 & 0.25 &   6.48 & 0.14 \nl
 7235 & [Ar IV]          &   1.60 & 0.06 &   0.90 & 0.03 \nl
 7254 & O I              &   0.23 & 0.06 &   0.13 & 0.03 \nl
 7281 & He I             &   1.92 & 0.06 &   1.07 & 0.03 \nl
 7320 & [O II]           &  11.23 & 0.08 &   6.24 & 0.05 \nl
 7330 & [O II]           &  12.68 & 0.08 &   7.03 & 0.05 \nl
 7749 & [ArIII]          &   2.34 & 0.06 &   1.22 & 0.03 \nl
 7818 &  He I            &   0.50 & 0.05 &   0.26 & 0.03 \nl
\enddata
\end{deluxetable}
     
\begin{deluxetable}{llrcrc}
\tablecolumns{6}
\tablewidth{0pc}
\tablecaption{Observed and reddening corrected spectrum of Wray~16-423 \label{tbl5}}

\tablehead{
\colhead{$\lambda$ (\AA)} & \colhead{IDENTIFICATION}  & \multicolumn{2}{c}{OBSERVED} &
\multicolumn{2}{c}{DEREDDENED} \\
\colhead{}    & \colhead{} & \colhead{FLUX}  & \colhead{ERROR} & \colhead{FLUX} & 
\colhead{ERROR} 
} 
\startdata
 3686 & H I (B19)        &   0.57 & 0.27 &   0.65 & 0.30 \nl
 3691 & H I (B18)        &   0.72 & 0.29 &   0.82 & 0.32 \nl
 3697 & H I (B17)        &   1.02 & 0.25 &   1.15 & 0.28 \nl
 3704 & H I (B16)        &   1.75 & 0.30 &   1.97 & 0.34 \nl
 3712 & H I (B15)        &   1.86 & 0.30 &   2.09 & 0.34 \nl
 3723 & H I (B14)        &   4.20 & 0.26 &   4.73 & 0.29 \nl
 3726 & [O II]           &  16.55 & 0.25 &  18.62 & 0.29 \nl
 3728 & [O II]           &   7.86 & 0.24 &   8.84 & 0.27 \nl
 3734 & H I (B13)        &   2.44 & 0.28 &   2.74 & 0.32 \nl
 3750 & H I (B12)        &   2.68 & 0.33 &   3.02 & 0.37 \nl
 3754 & O III            &   0.86 & 0.37 &   0.96 & 0.41 \nl
 3760 & O III            &   1.00 & 0.25 &   1.12 & 0.28 \nl
 3770 & H I (B11)        &   3.71 & 0.25 &   4.16 & 0.28 \nl
 3798 & H I (B10)        &   4.50 & 0.27 &   5.03 & 0.30 \nl
 3819 & He I             &   1.14 & 0.23 &   1.27 & 0.25 \nl
 3835 & H I (B9)         &   6.24 & 0.25 &   6.96 & 0.28 \nl
 3868 & [Ne III]         &  66.73 & 0.67 &  74.20 & 0.74 \nl
 3889 & H I (B8) + He I  &  16.72 & 0.25 &  18.56 & 0.27 \nl
 3967 & [Ne III]         &  21.31 & 0.21 &  23.48 & 0.23 \nl
 3970 & H$\epsilon$ (B7)   &  12.14 & 0.16 &  13.37 & 0.18 \nl
 4026 & He I             &   2.62 & 0.21 &   2.88 & 0.23 \nl
 4069 & [S II]           &   1.93 & 0.24 &   2.10 & 0.26 \nl
 4076 & [S II]           &   0.61 & 0.17 &   0.66 & 0.18 \nl
 4101 & H$\delta$ (B6)     &  23.58 & 0.25 &  25.64 & 0.27 \nl
 4143 & He I             &   0.68 & 0.23 &   0.74 & 0.25 \nl
 4267 & C II             &   0.46 & 0.14 &   0.49 & 0.15 \nl
 4340 & H$\gamma$ (B5)     &  44.69 & 0.21 &  47.38 & 0.22 \nl
 4363 & [O III]          &  12.92 & 0.20 &  13.67 & 0.21 \nl
 4387 & He I             &   0.70 & 0.09 &   0.74 & 0.10 \nl
 4471 & He I             &   5.20 & 0.22 &   5.43 & 0.23 \nl
 4517 & C~III?           &   3.39 & 0.18 &   3.52 & 0.19 \nl
 4541 & He II            &   0.31 & 0.07 &   0.32 & 0.08 \nl
 4569 & [Mg I]            &   0.30 & 0.09 &   0.31 & 0.10 \nl
 4685 & He II            &  11.14 & 0.13 &  11.36 & 0.13 \nl
 4712 & [Ar IV] + He I   &   2.99 & 0.08 &   3.04 & 0.09 \nl
 4740 & [Ar IV]          &   2.86 & 0.07 &   2.90 & 0.08 \nl
 4861 & H$\beta$ (B4)      & 100.00 & 0.00 & 100.00 & 0.00 \nl
 4922 & He I             &   1.34 & 0.07 &   1.33 & 0.07 \nl
 4959 & [O III]          & 377.21 & 0.99 & 373.07 & 0.98 \nl
 5007 & [O III]          & 1102.69 & 2.71 & 1084.79 & 2.67 \nl
 5048 & He I             &   0.22 & 0.06 &   0.22 & 0.06 \nl
 5412 & He II            &   1.03 & 0.06 &   0.97 & 0.05 \nl
 5518 & [Cl III]         &   0.27 & 0.05 &   0.25 & 0.04 \nl
 5538 & [Cl III]         &   0.37 & 0.04 &   0.34 & 0.04 \nl
 5755 & [N II]           &   0.58 & 0.06 &   0.53 & 0.06 \nl
 5795 & He II             &   0.40 & 0.03 &   0.37 & 0.03 \nl
 5814 & He II            &   0.20 & 0.04 &   0.19 & 0.04 \nl
 5876 & He I             &  18.07 & 0.10 &  16.37 & 0.09 \nl
 6103 & [K IV]           &   0.15 & 0.04 &   0.14 & 0.04 \nl
 6302 & [O I]            &   2.41 & 0.07 &   2.12 & 0.06 \nl
 6313 & [S III]          &   2.05 & 0.07 &   1.80 & 0.06 \nl
 6366 & [O I]            &   0.94 & 0.06 &   0.83 & 0.05 \nl
 6550 & [N II]           &   6.30 & 0.06 &   5.44 & 0.05 \nl
 6563 & H$\alpha$ (B3)   & 329.14 & 0.83 & 284.02 & 0.72 \nl
 6586 & [N II]           &  20.58 & 0.09 &  17.73 & 0.08 \nl
 6680 & He I             &   4.66 & 0.20 &   3.99 & 0.17 \nl
 6719 & [S II]           &   1.79 & 0.10 &   1.53 & 0.09 \nl
 6733 & [S II]           &   3.36 & 0.09 &   2.86 & 0.07 \nl
 7065 & He I             &   8.74 & 0.10 &   7.31 & 0.08 \nl
 7138 & [Ar III]         &  11.71 & 0.12 &   9.76 & 0.10 \nl
 7236 & [Ar IV]          &   0.44 & 0.05 &   0.37 & 0.04 \nl
 7263 & [Ar IV]          &   0.12 & 0.04 &   0.10 & 0.03 \nl
 7281 & He I             &   0.99 & 0.08 &   0.82 & 0.06 \nl
 7321 & [O II]           &   2.50 & 0.06 &   2.06 & 0.05 \nl
 7331 & [O II]           &   3.18 & 0.06 &   2.62 & 0.05 \nl
 7750 & [Ar III]         &   2.29 & 0.05 &   1.85 & 0.04 \nl
\enddata
\end{deluxetable}
     
\begin{figure}
\epsscale{0.80}
\plotone{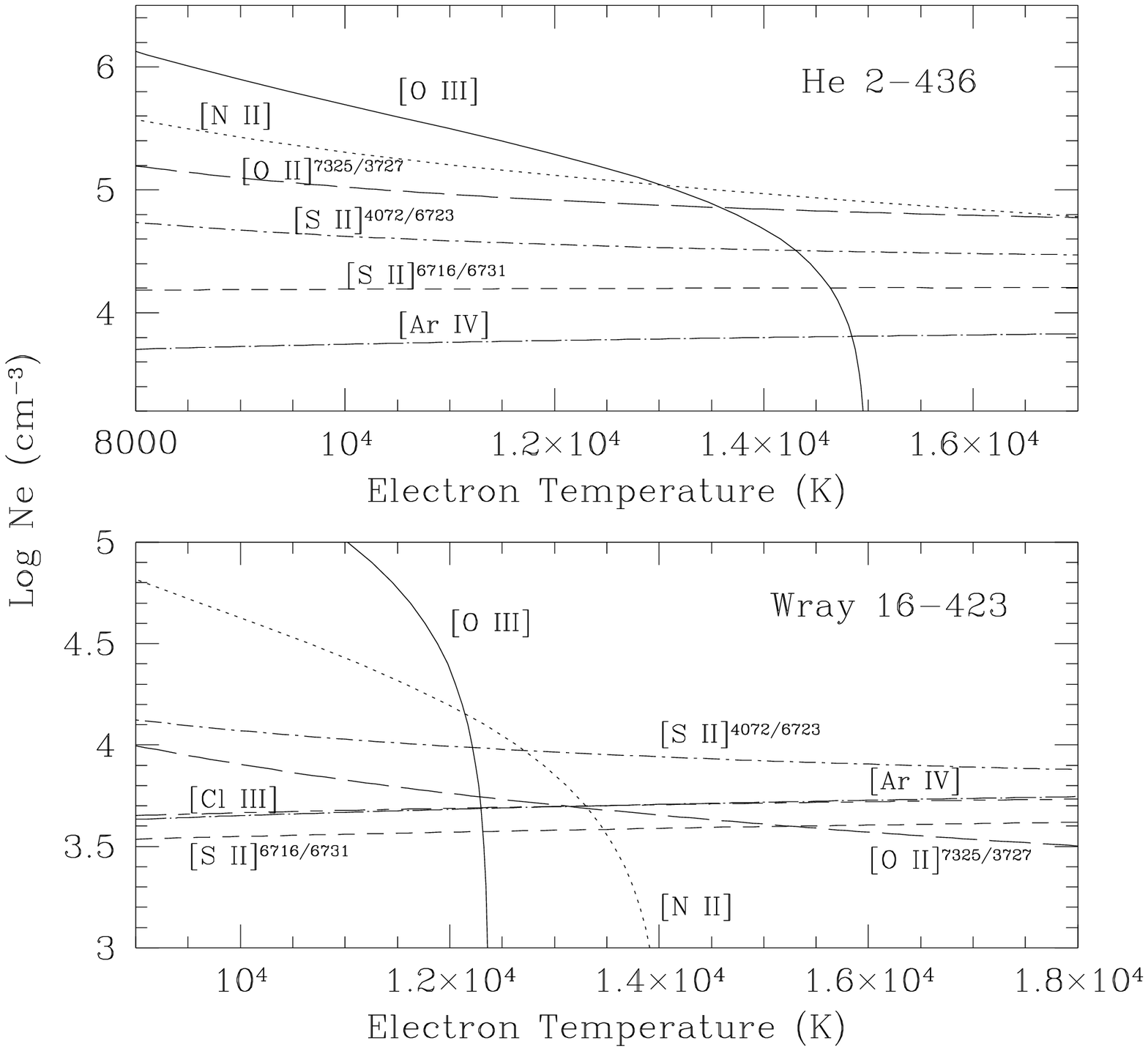}
\caption[]{N$_e$ - T$_e$ diagnostic diagrams for He~2-436 and Wray~16-423
are shown. The line ratios from which the diagnostic curves are derived
are indicated. For ratios
not indicated the following holds: [O~III] = 5007/4363\AA; [N~II] = 6583/5755\AA;
[Ar~IV] = 4711/4740\AA; [Cl~III]5517/5537\AA.\label{fig2}}
\end{figure}

 In Table 7 the observed and dereddened spectra of the Fornax PN are 
presented. The flux measurements are not of high quality; the extinction
is zero to within the error and the [O~III]4363\AA\ line was not
detected, nor were the [S~II]6716+6731\AA\ lines. The 
relative fluxes are, within the errors, generally consistent with those 
presented by Danziger et al. (1978) with the exception of the [N~II]
6583\AA\ line which is 35\% stronger (compared to [O~II]3727\AA) than
measured by Danziger et al. (1978) based on deblending of
the [N~II] line from H$\alpha$. Until improved data are
available, and in particular a determination of the electron density,
the abundances presented by Maran et al. (1984), based on the Danziger et al. 
(1978) fluxes, will be used, with exception of the N abundance which is 
increased by 35\%.

\begin{table*}
\begin{center}
\caption{Physical parameters of Sagittarius PN \label{tbl6}}
\begin{tabular}{lrrrr}
 Parameter & \multicolumn{2}{c}{He~2-436} & 
\multicolumn{2}{c}{Wray~16-423} \\
            & Value & Error & Value & Error \\
\tableline

c & 0.61 & 0.05 & 0.20 & 0.04 \\
E$_{B-V}$ (mag.) & 0.42 & 0.03 & 0.14 & 0.03 \\
Log$_{10}$F(H$\beta$)\tablenotemark{a} & -12.17\tablenotemark{b} & 0.03 & -12.09 & 0.03 \\ 
Log$_{10}$F(H$\beta$)\tablenotemark{c} & -11.56 & 0.06 & -11.89 & 0.04 \\ 
T$_e$ (K) & 12600 & 400 & 12400 & 400 \\
N$_e$ (cm$^{-3}$) & 1.1$\times$10$^{5}$ & 0.3$\times$10$^{5}$ & 6000 & 1500 \\
Ionized mass (M$_{\odot}$) & 0.016 & & 0.14 & \\
T$_{Zanstra}$ (H~I) & 61000 & & 85000 & \\
T$_{Zanstra}$ (He~II) & & & 96000 & \\
B (mag.)\tablenotemark{d} & 16.7 & & 18.7 & \\
L$_{\odot}$ & 5300 & & 2300 & \\
\end{tabular}
\end{center}
\tablenotetext{a}{Observed flux}
\tablenotetext{b}{Absolute observed H$\beta$ flux and error from  Webster (1983)}
\tablenotetext{c}{Dereddened flux}
\tablenotetext{d}{Corrected for extinction}
\end{table*}

\begin{table*}
\begin{center}
\caption{Observed and redenning corrected spectrum of Fornax PN \label{tbl7}}
\begin{tabular}{llrrrr}
 ~$\lambda$ (\AA) & IDENTIFICATION  & \multicolumn{2}{c}{OBSERVED} &
\multicolumn{2}{c}{DEREDDENED} \\
    &                & FLUX~  & ERROR & FLUX~ & ERROR \\ 
\tableline
     
 3727 & [O II]           &   35~~ & 6~~~ & 34~~ & 6~~ \\ 
 3869 & [Ne III]         &   45~~ & 5~~~ & 43~~ & 4~~ \\ 
 3889 & H I (B8) + He I       &   22~~ & 4~~~  & 21~~ & 4~~ \\
 3970 & [Ne III] + H$\epsilon$ (B7)   &   25~~ & 4~~~  & 24~~ & 3~~ \\
 4101 & H$\delta$ (B6)       &   28~~ & 3~~~  & 27~~ & 3~~ \\
 4340 & H$\gamma$ (B5)       &   38~~ & 3~~~  & 37~~ & 3~~ \\
 4861 & H$\beta$ (B4)        &   100~~ & 0~~~ & 100~~ & 0~~ \\
 4959 & [O III]          &   179~~ & 7~~~ & 180~~ & 7~~ \\
 5007 & [O III]          &   548~~ & 19~~~ & 551~~ & 19~~ \\
 5876 & He I             &   12~~ & 2~~~ & 13~~ & 2~~ \\
 6563 & H$\alpha$  (B3)  &   271~~ & 21~~~ & 285~~ & 12~~ \\
 6583 & [N II]           &   12~~ & 3~~~ & 13~~ & 3~~ \\
\end{tabular}
\end{center}
\tablecomments{Reddening c=$-$0.07$\pm$0.13 (E$_{B-V}$=-0.04)}
\end{table*}

Figure 2 shows N$_e$ - T$_e$ diagnostic diagrams for He~2-436 and
Wray~16-423. For Wray~16-423 there is a fairly well defined intersection 
region between the [O~III]5007/4363\AA, [N~II]6583/5755\AA\ and 
[Ar~IV]4711/4740\AA, [Cl~III]5517/5537\AA, [S~II] and [O~II]
diagnostic ratios; taking account of the errors in the ratios,
values for T$_e$ of 12400K and N$_e$ of 6000cm$^{-3}$ were adopted.
For He~2-436, the [O~II] and [S~II] diagnotics shows densities close 
to the critical densities for the upper levels of various transitions.
The only reliable intersection is for the [O~III], [N~II] and
[O~II]7325/3727\AA\ ratio; the adopted values of T$_e$ and N$_e$ are
12600K and 1.1$\times$10$^{5}$ respectively. Estimated errors on T$_e$ 
and N$_e$ for both nebulae are listed in Table 6. Considering the
observational errors and their propagation to the derived diagnotic parameters,
it was not considered warranted to distinguish two electron temperature regimes
as measured from the [O~III] and [N~II] diagnostic ratios. 
Ionized masses of both nebulae were calculated from the absolute H$\beta$ fluxes 
and the adopted electron densities and are listed in Table 6.

Using the adopted diagnostics, the abundances of the elements were derived
for the forbidden line species from five level atoms calculations using the
NEBULA package (\cite{SH95}). For He~I the recent recombination 
coefficients of Smits (1996) were used and for He~II the values tabulated 
by Osterbrock (1989). Using standard ionization correction factors 
(\cite{BA83} and \cite{KB94}), the total abundances of He, O, N, Ne, S, Ar 
were derived and are listed in Table 8 with the maximum one-sided errors 
(based on propagated errors in both T$_e$ and N$_e$ and the line ratios). 
The total C abundance is not listed in Table 8, since
only the C~II 4267\AA\ recombination line was observed; this line is
notoriously unreliable for C/H determination (see \cite{RS95}). 
The 4267\AA\ line was detected in both nebulae (signal-to-noise $>$3)
and in multiple spectra. The C$^{2+}$/O$^{2+}$ ratio ($\approx$ C/O) is 
7.8$\pm$2.3 for He~2-436 and 3.4$\pm$1.2 for Wray~16-423 (errors from
the C~II/[O~III] line ratio), indicating that both PN may be carbon rich.
This would not be surprizing given that there are carbon stars observed
in the Sagittarius galaxy (\cite{WHI96}) and a carbon Mira has been
discovered (Whitelock, 1997, private communication). Maran et al. (1984)
determined that the PN in Fornax is also carbon rich
(C/O=3.7), using a measurement of the forbidden C~III] doublet at 1908\AA.

\subsection{Imaging}
 For both the [O~III] and H$\alpha$ images, a bright unsaturated star image
was selected and used as the PSF for a Richardson-Lucy restoration
of the region around the PN images. 100 iterations of the Lucy (1974)
iterative deconvolution were performed. The image of He~2-436 appeared stellar 
(i.e. unresolved) in both filter images, implying a diameter $<$1$''$. 
For Wray~16-423, the source appeared to be elliptical with a major axis
of 1.2$''$ and axial ratio 0.7 in position angle $\sim$ 50$^\circ$. Adopting 
another star in the field for the PSF did not alter these results
significantly. Higher resolution (e.g. HST) images are required to resolve 
both nebulae and give more information on their morphology.

\section{Discussion}

\subsection{Abundances of PN in dwarf elliptical galaxies}
Table 8 contrasts the abundances of the light elements measured in the two
PN in the Sagittarius dwarf elliptical with that in the Fornax dwarf galaxy 
(from \cite{MAR84}). The similarity in abundances of O, Ne, S and Ar of 
the two PN in the Sagittarius dwarf 
can be regarded as independent evidence, in addition to the dynamical data, 
that both Wray~16-423 and He~2-436 do belong to the Sagittarius Dwarf galaxy. 
The abundance values listed for Fornax in Table 8 must be considered more uncertain 
in view of the lack of a direct measurement of the electron density
in this nebula. The striking aspect is that the O, Ne and Ar abundances
are identical within the errors between the Fornax PN and the Sagittarius PN.
This level of similarity strongly suggests
that all three PN were formed within a common evolutionary scenario and, since they
arise in similar mass stellar systems, at similar evolutionary times. The largest
discrepancy between these three PN is for N, where the N abundance of
He~2-436 is very low. However the ionization correction factor is large
for this element and the diagnostic diagram shows a range of physical conditions;
a lower electron temperature and or a higher electron density in the low 
ionization zone would increase the N abundance. The abundances
are much lower (by 0.5 dex on average) than Galactic PN as indicated by the 
comparison in Table 8 with the mean abundance for 42 non-Type I (i.e. from low 
mass progenitors) Galactic PN from Kingsburgh \& Barlow (1994). 
Solar abundances are included for comparison in Table 8 (column 9).

  Confirmation that He~2-436 and Wray~16-423 do belong to the Sagittarius
dwarf can be sought by comparing the nebular abundances with those of the 
stellar population.
The PN abundances give a mean depletion of -0.7 with respect to Solar ([X/H]). 
However in the case of the stars there appears to be a wide range of 
metallicities so direct comparison of PN and stellar metallicities is
equivocal. Sarajedini \& Layden (1995)  found two populations at
[Fe/H] = -0.52 and -1.2, as well as a more metal-poor population 
associated with M~54; Mateo et al. (1995) found a mean
metallicity of [Fe/H] = -1.1 and Whitelock et al. (1996) a mean value of 
[Fe/H] = -0.8. Marconi et al. (1997) observed a spread in [Fe/H] between 
-0.7 and -1.58, with a dominant population similar to the cluster 
Terzan~7 and Ibata et al. (1997) measured a metallicity range between [Fe/H] =
-0.8 and -1.0. It can be concluded that the Sagittarius PN progenitors are 
representative of the higher metallicity tail of the Sagittarius population. 
Choosing as compromise a mean [Fe/H] of -0.8, leads to a relative oxygen 
abundance of [O/Fe] = +0.2 for Sagittarius. 

It is natural to compare the abundances of the Sagittarius and Fornax PN 
with those of the `halo' PN. The halo PN are not a well defined group, 
being selected on the basis of either position away from the Galactic plane, 
high velocities and or low abundances. Their abundances
vary over a rather wide range so it is not appropriate to form a
mean abundance set (see Howard et al (1997) for an abundance analysis of
9 halo PN). Clegg et al. (1987) compared the abundances of three halo
objects H~4-1 (PN G 049.3 $+$88.1), BoBn-1 (PN G 108.4 -76.1) and DdDm-1 
(PN G 061.9 $+$41.3). It is clear that the abundances of the
Fornax and Sagittarius PN closely resemble those of DdDm-1 to a remarkable degree,
and this PN is included in Table 8 (data from \cite{CL87}). The possible
candidate PN in Sagittarius, PRMG-1 (PN G 006.0 $-$41.9), is included in 
Table 8 although the N and S abundances are limits only; the abundances 
are also similar to the Sagittarius and Fornax PN.  
Barker \& Cudworth (1984) have suggested that S and Ar, which should not 
be affected by the AGB nuclear processing, should be representative of 
the heavy element abundance of the parent material. The similarity in abundances 
of the Sagittarius and Fornax PN and DdDm-1 in these elements confirm their 
origin in a population which did not undergo large element enrichment from supernova 
products. In contrast the abundances of all elements are much higher, 
by $\sim$0.8dex, than for K~648, the PN in the metal poor Galactic globular 
cluster M~15 (\cite{AD84}), which is also listed in Table 8.

 In PN with low mass progenitors (non Type I), the abundance of O is 
only marginally affected by AGB evolution ($\leq$0.2dex, e.g. \cite{KB94}). 
However, the sum of C, N and O should change little through the AGB evolution. 
The increase in carbon is due to triple-$\alpha$ reactions,
the products of which are brought to the surface by the 3rd dredge-up. 
Assuming that the derived C/O abundances from the C~II 4267\AA\ line are
fairly reliable for the Sagittarius PN, both show excess carbon indicating that
He shell flashes have occured. The C abundance of the Fornax PN is more reliable
and shows a similar level at C/H of 1$\times$10$^{-3}$. In contrast DdDm-1 shows
low carbon abundance (12+Log(C/H)$<$7.1, \cite{CL87}) and it has been suggested
that most of the C was converted into N and that mass loss of the low mass core 
terminated after only early AGB evolution (\cite{CL87}). The high carbon abundance
in the dwarf galaxy PN suggest that these stars have progressed through a number of
thermal pulses and suffered He shell burning. The lower N abundance for He~2-436 
could be due to relatively little carbon being burnt to N (the 12+Log(C/H) 
abundance of He~2-436 is 9.2, whilst it is 8.8 for Wray~16-423, taking the C/O 
ratios at face value).

  The fundamental question to be answered is: what do the abundances of the PN
in the Sagittarius and Fornax dwarf galaxies tell us about the progenitor stars
from which they were formed, and further, what do they reveal of the
evolutionary history of these galaxies. The similarity of S and Ar abundances 
with the SMC PN abundances (\cite{MBC88}) is striking and, together 
with the proposal that the Sagittarius 
galaxy is quite massive (\cite{IBA97}), suggests that the difference between 
the SMC and Sagittarius can partly be understood in terms of gas being stripped 
from Sagittarius and with little interaction occurring with gas clouds, in 
contrast to the SMC. The fact that the PN
abundances are almost as low as in the more metal rich blue compact galaxies
(such as II~Zw~40, \cite{WR93} and NGC~2363, \cite{PPPP86}) suggests that 
these satellite galaxies could be related objects with a different 
history of gas infall. If the elevated oxygen abundance of the
Sagittarius and Fornax PN relative to K~648 in M~15 is a guide,
then the Sagittarius and Fornax systems are not simply agglomerations of 
Galactic globular clusters. The similarity between the abundances of the two 
Sagittarius PN and those of PRMG-1 (\cite{PRMG89}), suggests that the latter 
PN could also belong to the Sagittarius dwarf. 
If PRMG-1 does belong to the Sagittarius galaxy and is still bound, 
then the extent of the galaxy is much larger than found by Ibata et al. (1997).
DdDm-1 also has similar abundances, but due to its location it would be 
premature to suggest that it was once a member of the Sagittarius dwarf. 
The similarity of abundances 
between the Sagittarius and Fornax PN and some halo PN suggests
that some halo stars could have been been captured 
from satellite dwarf galaxies such as Sagittarius and Fornax. 

The [O/Fe] value for the Sagittarius dwarf has implications for the metal 
enrichment history of this galaxy. An [O/Fe] value significantly higher 
than Solar would rule out Sagittarius as a dwarf irregular galaxy, but
taking account of the spread in [Fe/H] deduced from the various published 
determinations, this argument is weakened.  The presence of a wide metallicity range
indicates star formation at different times. Several pieces of evidence
point towards the notion that Sagittarius was more massive in the past:
it falls below the luminosity-metallicity relation of Lee (1995);  
it is being disrupted by the Milky Way (\cite{PIA95}; \cite{VW95};
\cite{JOH95}); it may have had an escape
velocity high enough to retain the gas and form successive generations
of stars (\cite{IBA97}). The fact that [O/Fe] is $\sim$ $+$0.2 would 
argue against significant enrichment from SN type I. In an isolated
dwarf galaxy, this value would favour a
SN driven wind followed by recapture of the enriched gas with subsequent
star formation. Thus, the available element ratios are not inconsistent
with two bursts of star formation, although the star formation history
of Sagittarius must be unusual. In particular, this history may
not be directly compared to that of other more distant
dwarfs that have not suffered severe interactions with the Milky Way.
The interation with the Galaxy may have caused the gas to escape after
a few orbits, and it may have caused prior bursts of star formation.

\begin{deluxetable}{lrrrrrrrr}
\tablecolumns{9}
\tablewidth{0pc}
\tablecaption{Comparison of abundances for dwarf elliptical galaxy and Galactic 
PN \label{tbl8}}
\tablehead{
\colhead{Element\tablenotemark{t}} & \colhead{He~2-436} & \colhead{Wray~16-423} & 
\colhead{Fornax\tablenotemark{u}} & \colhead{DdDm-1\tablenotemark{v}} &
\colhead{PRMG-1\tablenotemark{w}} & \colhead{K~648\tablenotemark{x}} &
\colhead{Galactic} & \colhead{Solar\tablenotemark{z}} \\
\colhead{} & \colhead{} & \colhead{} & \colhead{PN}  & \colhead{} & \colhead{} &
\colhead{} & \colhead{PN\tablenotemark{y}} &  \colhead{} \\
}
\startdata
He      & 11.02$\pm$0.02  & 11.03$\pm$0.02 & 11.08 & 11.0 & 10.96 & 11.02 & 11.05 & 10.99 \\
N       & 6.97$\pm$0.16 & 7.62$\pm$0.10 & 7.5 & 7.4 & $<$8.3 & 6.5 & 8.14 & 8.00 \\
O       & 8.29$\pm$0.08 & 8.31$\pm$0.07 & 8.4 & 8.1 & 8.06 & 7.67 & 8.69 & 8.93 \\
Ne      & 7.57$\pm$0.08 & 7.50$\pm$0.08 & 7.6 & 7.3 & 7.23 & 6.7 & 8.10 & 8.09 \\
S       & 6.30$\pm$0.08 & 6.48$\pm$0.08 &      & 6.5 & $<$7.00 & 5.2 & 6.91 & 7.24 \\
Ar      & 5.76$\pm$0.12 & 5.88$\pm$0.08 & 5.9 & 5.8 & 5.50 & 4.3 & 6.38 & 6.57 \\
\enddata
\tablenotetext{t}{12 + Log$_{10}$(Abundance/H)}
\tablenotetext{u}{Fornax PN O, Ne and Ar abundances from Maran et al. (1984)}
\tablenotetext{v}{DdDm-1 abundances from Clegg et al. (1987)}
\tablenotetext{w}{PRMG-1 abundances from Howard et al. (1997)}
\tablenotetext{x}{K~648 abundances from Adams et al. (1984), except for S and Ar
from Barker \& Cudworth (1984)}
\tablenotetext{y}{Mean composition for 42 non-Type I Galactic PN from 
Kingsburgh \& Barlow (1994)}
\tablenotetext{z}{Solar abundances from Anders \& Grevesse (1989)}
\end{deluxetable}

\subsection{The central stars of the Sagittarius PN}

Both He~2-436 and Wray~16-423 show Wolf-Rayet (WR) features in their spectra.  
Figure 3 shows part of the low dispersion ESO 1.5m spectra at the flux levels of
the broad stellar lines.  Table 9 lists the observed equivalent widths, line
widths (FWHM corrected for instrumental width) and fluxes: He~2-436 shows  
strong C~IV 4658 and 5810\AA\ and weaker N~III 4640\AA\ and He~II 4686\AA,
but no C~III lines (4650 and 5695\AA) were detected; 
Wray~16-423 shows weak N~III 4640\AA,  C~III 4650\AA\ and C~IV 
5810\AA, barely detected C~III 5695\AA, but no detectable broad He~II 
4686\AA\ emission. On the basis
of the relative line strengths He~2-436 is classified as type [WC3-4] since it
has strong C~IV, lack of C~III  and no O~V or O~VI; 
Wray~16-423 is classified as type [WC8] on the basis of the similar strengths of 
the C~III and C~IV lines and absence of O~V (c.f. the WR 
classification of PN by \cite{TASG93}). However this classification goes 
in the opposite sense to that expected from the Zanstra temperatures of both
stars (Table 6): He~2-436 has an earlier [WC] class but a lower temperature,
whilst Wray~16-423 has a later spectral class but considerably higher
stellar temperature. The weak and narrower lines of Wray~16-423 would lead  
to its classification as a ``weak emission line central star'' by
Tylenda et al. (1993), so it may be somewhat misleading to directly compare both
stars. However the discrepancy is puzzling when considered simply in terms
of ionization - the hotter star appears to produce a lower ionization
wind than the cooler star. This effect remains unexplained but could
possibly be due to different line blanketing in the two stars.

\begin{figure}
\plotone{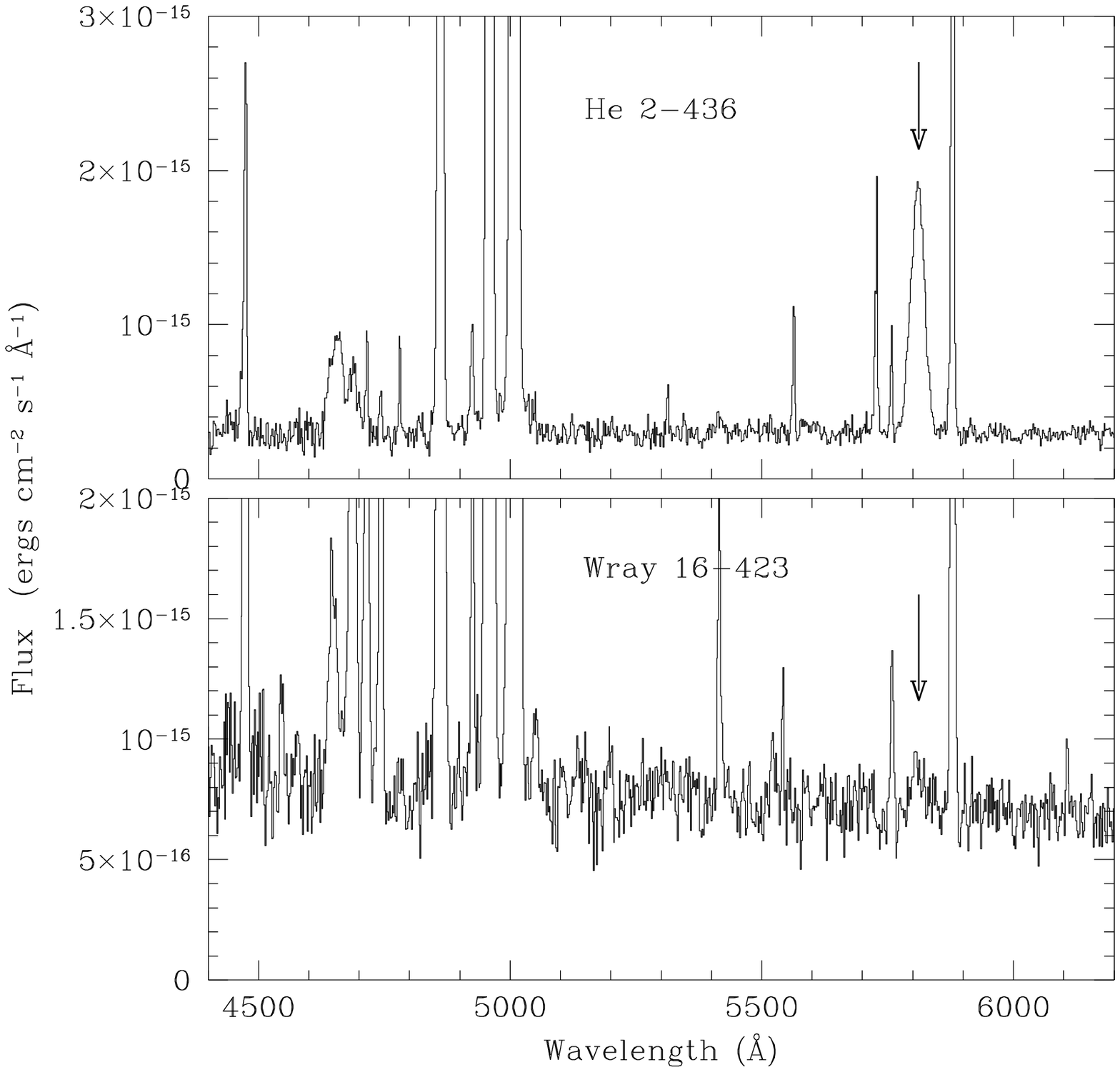}
\caption{Detail of the continuum spectra of He~2-436 and Wray~16-423 is
shown. The broad stellar lines of C~IV 5810\AA\ are arrowed and the
complex of broad lines from C~III, C~IV, N~III and He~II is visible around 4600\AA,
although weaker in Wray~16-423 where they are confused by the strong
nebular lines.\label{fig3}}
\end{figure}

\begin{table*}
\caption[]{Parameters of stellar emission lines in He~2-436 and Wray~16-423
\label{tbl9}}
\begin{center}
\begin{tabular}{lrlrrr}
 Target & ~$\lambda$(\AA) & IDENT. & EW(\AA) & FWHM(\AA) & Flux\tablenotemark{x} \\
\tableline
     
 He~2-436 & 4640 & N~III  & -15.3 & 12.3 & 3.7 \\
          & 4658 & C~IV   & -55.4 & 23.0 & 13.4 \\
          & 4686 & He~II  & -23.7 & 13.4 & 5.7 \\
          & 5810 & C~IV   & -164.6 & 31.1 & 32.9 \\
          &      &              &     &      &      \\
 Wray~16-423 & 4640 & N~III  & -12.7 & 13.1 & 1.0 \\
             & 4650 & C~III  & -5.4 & 9.9 & 0.45 \\
             & 5810 & C~IV   & -5.8 & 14.7 & 0.42 \\

\end{tabular}
\end{center}
\tablenotetext{x}{Observed flux in ergs cm$^{-2}$s$^{-1}$ $\times$10$^{-14}$.}
\end{table*}

The Wolf-Rayet characteristics of the two central stars give important
constraints on their evolutionary status. Central stars of planetary 
nebulae (CSPN) mostly evolve either on
hydrogen-burning (\cite{SCH83}) tracks or on helium-burning tracks,
before entering the white-dwarf cooling phase. The track for a given
object depends on the phase of the thermal-pulse cycle at which the
central star leaves the AGB. Since the
hydrogen-burning phase lasts several times longer than the
helium-burning one, most CSPN are assumed to belong to the former
category. In contrast, since the Wolf-Rayet CSPN are known to be
hydrogen poor, there is little doubt that they are presently in a
helium-burning phase (e.g. \cite{POT96}).
The wind emission lines, which are the defining characteristics of
WR stars, are expected to disappear on the white-dwarf
cooling track: the luminosity quickly drops by a factor of 100,
far below the Eddington luminosity. Consequently, [WC] stars will
still be on the nuclear-burning part of their evolution.  

Compared to the Sch\"{o}nberner tracks (\cite{SCH83}), which assume
quiescent hydrogen burning, helium burners are expected to deviate in
their evolutionary time scales and luminosity evolution  (\cite{IB84}).
Detailed calculations of evolutionary tracks including
helium-burning phases have been made by Vassiliadis \& Wood (1993) but
these are not yet comprehensive. Whereas quiescent hydrogen burners
are described by the core mass as the single major free parameter, helium
burners need in addition the phase of the thermal-pulse cycle which
measures where on the evolution the thermal pulse occured. In
addition, the models assume the presence of an inert hydrogen layer
above the helium layers. In the [WC] stars this layer is absent.  As
a consequence they are not expected to be well described by present
models (\cite{SCH96}, \cite{SCH97}) and extreme care has to be taken in
using helium tracks to describe Wolf-Rayet central stars. One of the
most favored scenarios for forming hydrogen-deficient central stars is
the so-called ``very late'' thermal pulse, whereby the
helium-flash starts after the cessation of hydrogen-burning (i.e. when
the star is on the white dwarf cooling track). The convective mixing that later
occurs transports any remaining hydrogen downwards into hotter layers
where it is burned. The C and O surface abundances predicted in this
scenario do not however agree with the observed ones (\cite{SCH96}). Evidence 
for the presence of helium-burning central stars among planetary nebulae 
in the LMC has been presented by Dopita et al. (1997).

Helium-burning central stars are expected to evolve on tracks
of decreasing luminosity towards higher stellar temperatures. The
final thermal pulse can occur at the tip of the AGB or during the post-AGB
evolution.  Immediately following the thermal pulse, the evolution is
very fast (\cite{VW93}), and stars in an early phase
after the thermal pulse may show up as bright infrared sources
(\cite{ZIL94}). Subsequently, during quiescent helium burning,
they evolve about three times slower than hydrogen-burning stars of
the same core mass. However the Wolf-Rayet wind has little effect
on the chemical composition of the nebula: during the time of heaviest
mass loss following the thermal pulse, the mixing which causes the
disappearance of the hydrogen envelope has not yet begun.  Galactic
planetary nebulae surrounding [WC] stars do not show differences in
their abundance patterns (\cite{POT96}, \cite{GOR95}).

The stellar luminosities of both PN are listed in Table 6 based on the
magnitudes and black body temperatures of the central stars. An estimate 
of the error on the derived luminosities is 20\%, resulting from
the errors on the extinction, the magnitude and the black body temperature.
For He~2-436, a lower limit to the luminosity can be derived by integration 
of the 12 and 25$\mu$m fluxes and the upper limits to the 60 and 100$\mu$m 
IRAS fluxes; a value of 4600L$_\odot$ is derived, consistent with the value
in Table 6. Helium-burning central stars can explain why the two Sagittarius 
PN have luminosities which differ by a factor of 2 (Table 6). The models for
helium-burning central stars (assuming an inert upper layer of
hydrogen) show that their luminosity along an evolutionary track
decreases by upto a factor of 0.6 dex, whereas hydrogen-burning
central star models keep a constant luminosity before entering their 
cooling track. The two PN are expected
to have similar core masses: for progenitor masses of 1 M$_\odot$ and
lower, the final mass is strongly peaked around a value of
approximately 0.55 M$_\odot$ with very small dispersion (\cite{WEI87}).
Only for stars of twice solar and higher masses are significantly higher
final masses expected. For hydrogen-burning tracks this would lead to
similar luminosities as well, excluding the white-dwarf cooling
track. The different luminosities indicate that at least one (the one
with the lower luminosity) is not evolving on a track of quiescent
hydrogen burning. The position on the HR diagram indicates that
neither star has entered the white-dwarf cooling track. 

He~2-436 has a higher luminosity than Wray~16-423 and the central star
is cooler. The nebula is still young, as indicated by its high density, 
low ionized mass (Table 6), compact angular
size (unresolved in our images) and strong infrared emission; 
it has shown fast evolution from the AGB. One can hence infer that the
star is in an early post-helium-flash evolution. Its luminosity
should still be close to the corresponding hydrogen-burning
phase. Assuming therefore the hydrogen core mass--luminosity relation
of Vassiliadis \& Wood (1993), a core mass of 0.59 M$_\odot$ is indicated. 
Wray~16-423 is more evolved, based on the lower luminosity, lower density and 
larger angular size (the nebula is resolved with a diameter of 1.2$''$), as
well as lack of infrared emission. Assuming an expansion velocity of 15
km s$^{-1}$ (\cite{ZW96}) and using the distance of 25 kpc to the Sagittarius
dwarf galaxy, a dynamical age of 4000 years can be deduced for Wray~16-423. 
Competing factors contribute to the large uncertainty that should be assigned to
this figure. As the PN shell enters its energy-driven phase (\cite{KAH90}),
the pressure due to the hot shocked gas filling the inner cavity can accelerate the
shell. On the other hand, the outer parts of the shell trace the
material expelled during the early superwind phase when the wind
velocity might have been lower: the details of the
velocity law along the superwind phase are not known. 

The models of Vassiliadis \& Wood (1993)
or Bl\"{o}cker (1995a, 1995b) give estimates for
the transition time from the tip of the AGB to the post-AGB phase and
then for the evolutionary time scale on their tracks. The uncertainty on the
former quantity is quite large due to different definitions of the end
of the AGB between these authors and even changes quite substantially
between the models. A dynamical age of 4000 years is still consistent
with the time scale needed by a 1 M$_\odot$ star to evolve from the top of
the AGB to the actual position of the Sagittarius PN central stars on a 
log L--log T$_{eff}$ diagram. Provided that the comparison to helium-burning
tracks is relevant one can conclude that although no single track in 
Vassiliadis \& Wood (1993) can be uniquely identified with both PN,
observationally the results are consistent with Wray~16-423 and He~2-436 evolving
on the same track. 

Wolf-Rayet central stars make up about 10-15\% of the Galactic PN population.
They are found both in the disk and bulge, but none are known in the
Galactic halo. This indicates a difference between halo/globular
cluster PN population (no [WC] stars among ten known PN) and Sagittarius (two
out of two). However the presence of weak emisssion line stars, such as
Wray~16-423, requires deep spectra and broad stellar features may have been missed.
This difference between halo PN and those in Sagittarius could be explained 
if the halo PN have
somewhat lower luminosities, and are therefore unable to sustain a wind.
The fact that, from stellar population studies (\cite{MAT96}), 
the bulk of the stellar population in Sagittarius appears 
younger than the Galactic halo is consistent with this. Metallicity might
also be a factor, however at this stage of evolution the stellar
abundances are dominated by self-produced metals.  
A possible explanation of the lack of [WC] stars in the halo is that the [WC]
phase at lower core mass could be more confined to the high-luminosity
peak following the thermal pulse where evolution is very fast. 

\subsection{Sagittarius as a dwarf elliptical galaxy}

The Sagittarius dwarf is now heavily distorted by the tidal field  of
the Milky Way (e.g. \cite{IBA97}, \cite{PIA95}), making it difficult to decide 
its original type. 
Even though it appears to have a substantial fraction of intermediate-age
stars (e.g. \cite{SAR95}, \cite{MAT96}),
the galaxy presently contains no neutral hydrogen (\cite{KO94}),
an argument favouring a dwarf elliptical over a dwarf irregular galaxy.
The gas could have been stripped by its many passages through
the plane of the Milky Way (\cite{IBA97}). 
Figure 4 shows the [O/Fe] abundance ratio plotted against the logarithmic
oxygen abundance for the Sagittarius dwarf, together with data for the
Milky Way and its satellite galaxies and for the dwarf satellites NGC~185 and
NGC~205 of M~31. In this figure the O abundances are derived from the PN
and the Fe from the stellar population. The data for NGC~185 and NGC~205 
are taken from Richer et al. (1997).
For NGC~205 the oxygen abundance is based on lower limits derived from
[O~III] and H$\beta$ line fluxes for 9 PN (\cite{RIC95}), whilst for
NGC~185 it is based on spectroscopy of two PN. The oxygen data for the 
Milky Way disk and bulge are from Ratag et al. (1992).

 For the Fornax dwarf, Richer \& McCall (1995) estimate an oxygen abundance 
0.4 dex lower than that measured for the PN (\cite{MAR84}; \cite{DAN78}) 
based on the fact that the oxygen 
abundance is slightly higher for the brightest PN in samples for the
LMC, SMC and the Galaxy. However it can be argued that if this is the only
PN in this galaxy, which is consistent with the mass of the galaxy (\cite{ZW96}), 
then such a correction is not applicable. A further argument against applying
such a correction is that for the two Sagittarius PN, one (He~2-436) is about 
twice the luminosity of the other whilst their oxygen abundances are very similar.
Inferring the oxygen abundance of a stellar population based on a few PN
is clearly a complex task and not one that can immediately be referred to
more massive galaxies where samples are large and the star formation
history may be quite different. The oxygen abundances for NGC~185 and NGC~205 
used in Figure 4 have been determined by applying a correction to the
lower limits based on the behaviour of Milky Way and Magellanic
Clouds PN samples. From Figure 4 it is clear that the [O/Fe] abundance of 
the Sagittarius galaxy is higher than that in the Magellanic Clouds by a 
factor of $\sim$ 2, which also argues against Sagittarius having been an 
SMC-like galaxy.

\begin{figure}
\plotone{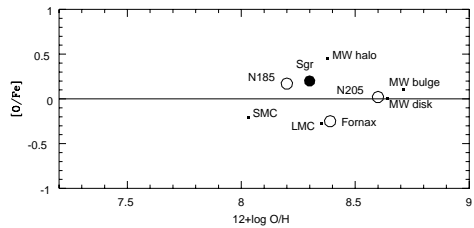}
\caption{Plot of the [O/Fe] abundance ratio versus the logarithmic
oxygen abundance (12.0 + Log$_{10}$(O/H)) for dwarf elliptical galaxies
around the Milky Way and M~31. The source of the data for the Milky Way
and the M31 satellite galaxies (NGC~185 and 205) is \cite{RIC95}. \label{fig4}}
\end{figure}

The Sagittarius PN have somewhat higher abundances than those in the Milky Way halo 
(see Table 8), typical of younger PN than the halo ones and much higher than
K~648. Mateo et al. (1995) argue that the
specific frequency of RR Lyrae variables in Sagittarius is low compared with other
dwarf elliptical galaxies, which is consistent with a higher metallicity and
relative youth of Sagittarius with respect to the halo. This indicates
that the halo could not have been made {\it entirely} of dwarf galaxies such 
as Sagittarius.

\begin{figure}
\plotone{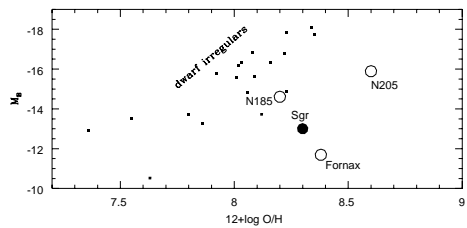}
\caption{Plot of the absolute blue magnitude of the satellite dwarf galaxies
of the Milky Way and M~31 against the oxygen abundance (12.0 + Log$_{10}$(O/H)).
The total magnitudes are taken from \cite{RIC95} as are the oxygen abundances
for the dwarfs except for Sagittarius (this work) and Fornax (oxygen abundance
from \cite{MAR84}). \label{fig5}}.
\end{figure}

The total absolute magnitude of Sagittarius is unknown. Based on the number of
RR Lyrae stars, Mateo et al. (1995) estimated a total magnitude $M_V = -13$,
and $-11$, if Sagittarius had a total of 1930 and 310 RR Lyrae (type ab) stars
respectively. Alard (1996) and Alcock et al. (1997) have
discovered more than 350 RR Lyrae type ab stars in the northern extension of 
Sagittarius alone.
Their surveys cover at most 1/5 of the total extension of Sgr, judging from
the isophotes, so it is concluded that Sagittarius contains $N > 1800$ RR Lyrae 
type ab stars. Following the arguments of Mateo et al. (1995), a value of $M_V = -13$ 
is adopted for Sagittarius, which is consistent with a minimum total mass of 
10$^9$M$_{\sun}$ (\cite{IBA97}). Sagittarius also has at least 4 globular 
clusters (\cite{IG95}) but there are a number of
globular clusters in the Galactic bulge region ($\approx$30)
without velocity or distance information  that could also be members.
The data accumulated so far (see in particular \cite{IBA97}) indicates that 
Sagittarius is at least as massive as Fornax, and it is likely that this 
galaxy is the most massive of all the dwarf satellite galaxies of the Milky Way. 

It is apparent that the oxygen abundance in dwarf elliptical 
galaxies is higher than for dwarf irregulars at the same absolute magnitude
(\cite{RIC95}), by about a factor of $\sim$2.
Figure 5 shows the dependence of the galaxy luminosity on the oxygen abundance
for all the Local Group dwarfs (irregulars and ellipticals) with measured PN
or H~II region O abundances. This figure shows that, even though the total 
absolute magnitude of Sagittarius
is still uncertain, this galaxy shares the location with the typical dwarf
ellipticals NGC~185, NGC~205, and Fornax. It is concluded that the
present study of the Sagittarius PN, added to the previous evidence,
supports the notion that Sagittarius was not a dwarf irregular,
but most probably a dwarf elliptical galaxy.

\section{Conclusions}
Optical spectra of the two planetary nebulae in the Sagittarius dwarf galaxy 
have been analysed and the abundances and physical conditions studied.
Both PN display similar abundances, differing considerably only for
nitrogen. The abundance pattern is also very similar to that for the PN
in the Fornax dwarf galaxy. The abundance of oxygen (for example) in these
three PN is lower than that of Galactic PN, but not as low as for genuine 
halo PN, such as the PN in the globular cluster M~15. Broad stellar (WR) 
lines are seen in both nebulae in Sagitarius with 
He~2-436 displaying the stronger lines. The luminosity of He~2-436 is
about twice that of Wray~16-423 and the latter has a resolved ionized shell.
In order to reconcile the different luminosities and evolutionary
characterisitics of He~2-423 and Wray~16-423, it is suggested that both
are on helium burning evolutionary tracks. The oxygen abundances of
the M31 and Milky Way satellite galaxies, as revealed by their planetary nebulae, 
are compared and the evidence suggests that both Sagittarius and Fornax were
not dwarf irregulars but more probably dwarf elliptical galaxies before 
their interaction with the Milky Way. 

\acknowledgements
The ESO staff at La Silla and the NTT remote observing team in Garching 
provided excellent support for the observations. We are very grateful to
the South African Astronomical Observatory for prompt retrieval of the 
spectroscopic data for the Fornax PN from their digital archive.
We would like to thank the referee, G. H. Jacoby, for many suggestions 
which substantially improved the paper. The work of D. Minniti was supported in 
part by Lawrence Livermore National Laboratory, under DOE contract W-7405-Eng-48.

\end{document}